# Comparison of multiple-gate MOSFET architectures using Monte Carlo simulation

Jérôme Saint-Martin, Arnaud Bournel, Philippe Dollfus

Institut d'Electronique Fondamentale, UMR CNRS 8622, Université Paris Sud, Bât. 220,
F-91405 Orsay cedex, France
Tel: (+33) 1 69 15 40 37, Fax: (+33) 1 69 15 40 20, E-mail: stmartin@ief.u-psud.fr

**Multiple-gate SOI MOSFETs with gate length equal to 25 nm are compared using device Monte Carlo simulation. In such architectures, the short channel effects may be controlled with much less stringent body and oxide thickness requirements than in single-gate MOSFET. Our results highlight that planar double-gate MOSFET is a good candidate to obtain both high current drive per unit-width and weak subthreshold leakage with large integration density and aggressive delay time, compared to non planar devices such as triple-gate or quadruple-gate structures.**

## Introduction

Multiple-gate structures on undoped SOI (Silicon On Insulator) are promising architectures likely to overcome short channel effects (SCE) in nanometer-scaled MOSFET [1]. Contrary to bulk MOSFETs, Double- (DG) [2], Triple- (TG) [3] and Quadruple-gate (QG) MOS transistors do not need drastic doping channel engineering. Moreover, they allow relaxing the oxide thickness $T_{ox}$ and the body thickness $T_{Si}$ requirements, which are severe in fully depleted Single Gate MOSFETs (SG) on SOI [4].

In this article, different intrinsic SOI MOSFET architectures, SG, DG, TG and QG, have been examined, in particular in terms of on-state $I_{on}$ and off-state $I_{off}$ currents. Considering nanometer scale devices where quasi ballistic transport is of great importance [5], simulations have been made using particle Monte Carlo method. SG and DG (resp. TG and QG) are simulated using a 2D (resp. 3D) Poisson solver. Details about the device Monte Carlo simulator may be found in Ref. [6]. Quantization effects are not taken into account. Scattering mechanisms with impurities, phonons and rough interfaces are considered in the algorithm.

## 1. Studied devices

Unless otherwise stated, the scaling of the considered structures obeys ITRS 2002 requirements corresponding to the high performance 65 nm-node [7]. The gate length $L_G$ (along x-axis) is 25 nm. The $SiO_2$ gate oxide $T_{ox}$ thickness is equal to the upper limit, 1.2 nm. The work function of the metallic gate material is 4.46 eV to achieve a threshold voltage $V_T$ of 0.2 V. The power supply voltage $V_{DD}$ is 0.7 V. The doping density is $N_D = 5 \times 10^{19}$ cm$^{-3}$ in $N^+$ source-drain regions and $N_A = 2 \times 10^{15}$ cm$^{-3}$ in the body (P type). Moreover, the $N^+/P$ junctions are assumed to be abrupt. In SG and TG, the buried oxide (BOX) thickness $T_{box}$ is 25 nm and the bulk doping (P type) is $2 \times 10^{17}$ cm$^{-3}$. The bulk is ground biased. The channel of TG and QG has a square cross section: the channel width W (along z-axis) between the side gates is equal to the body thickness $T_{Si}$. In the planar DG, the thickness $T_{Si}$ of the body (from $y = 0$ to $y = T_{Si}$) separates the top gate from the bottom gate.

In order to facilitate the comparison, the drain current $I_D$ per unit-width (A/m) is first calculated in all cases by dividing the drain current (A) resulting from Monte Carlo simulation by W. However different approaches may be used in current normalization [1], considering in particular the multi-fingered architecture of non planar devices (see Ref. [3] for TG). The intrinsic gate delay $C_G V_{DD}/I_{on}$, where $C_G$ is the gate capacitance at $V_{GS} = V_{DD}$ and low $V_{DS}$, may provide an unambiguous figure of merit to compare multiple-gate architecture.

In this work, ITRS 2002 recommendations will be used to provide acceptable performances but not strict specifications. We first consider reference devices in which $T_{Si}$ is 10 nm and the channel length $L_{ch}$ is 15 nm (the gate overlap on source-drain junctions is 5 nm and the length $L_{SD}$ of $N^+$ source-drain regions is 25 nm). Such low form factor $L_{ch}/T_{Si}$ and relatively high $T_{ox}$ are not favorable to control SCE in the SG device. Other paragraphs will analyze the influence of $T_{Si}$ and $L_{ch}$. At last, a discussion will summarize the main issues of multiple-gate scaling in terms of $C_G V_{DD}/I_{on}(I_{off})$ comparison.

## 2. Results

### 2.1 On-state characteristics

We first examine the electron transport in the simulated devices in the on-state ($V_{GS} = V_{DS} = V_{DD}$). As illustrated in Figure 1 that represents the percentage of electrons which have crossed the channel from source-end to drain-end as a function of the number of experienced scattering events $N_{int}$ (see Ref. [5]), the transport is strongly ballistic for these 15 nm long channels: the fraction of electrons decays exponentially as a function of $N_{int}$ in all devices, and the ballistic electrons are in a majority (about 50%). It is important to note that increasing the number of rough $SiO_2$/Si interfaces does not strongly impact the electron transport in the channel. Besides, similar curves are obtained for the thinner devices ($T_{Si}$ = 5 nm) studied. However the impact of the increase of rough interfaces is then more visible in the thinner QG as shown in Figure 1.

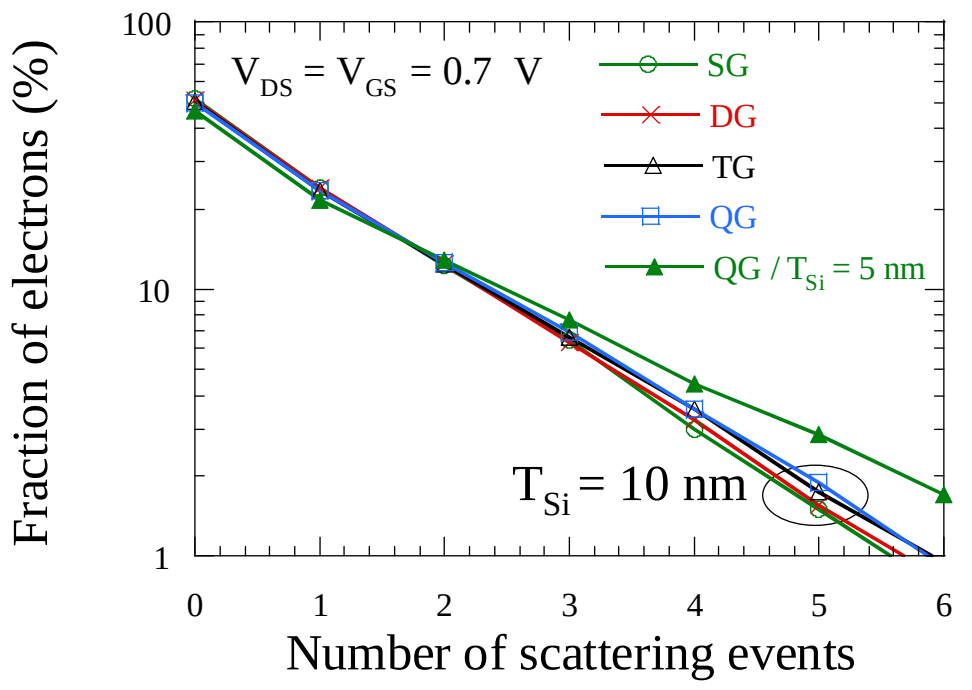

*Figure 1: Fraction of electrons flowing through the channel from source-end to drain-end versus the number of scattering events undergone in the on-state. Reference devices and thin QG ($T_{Si}$ = 5 nm).*

The $I_D(V_{DS})$ curves at $V_{GS} = V_{DD}$ and $I_D(V_{GS})$ curves at $V_{DS} = V_{DD}$ are plotted in solid lines for reference devices in Figure 2 and Figure 3, respectively. The on-state current $I_{on}$ in the SG (1625 A/m) is much higher than 2002 ITRS recommendations (900 A/m). However, other transistor characteristics are deplorable, for instance the drain conductance $g_D$ in the on-state is unacceptably high: 880 S/m. The current $I_{on}$, the transconductance $g_m$ at $V_{DS}$ = 0.7 V and $g_D$ are improved as the number of gates increases: 2140 A/m, 4170 S/m and 540 S/m respectively for the DG, 2420 A/m, 5700 S/m and 480 S/m for the TG and 2815 A/m, 7070 S/m and 370 S/m for the QG.

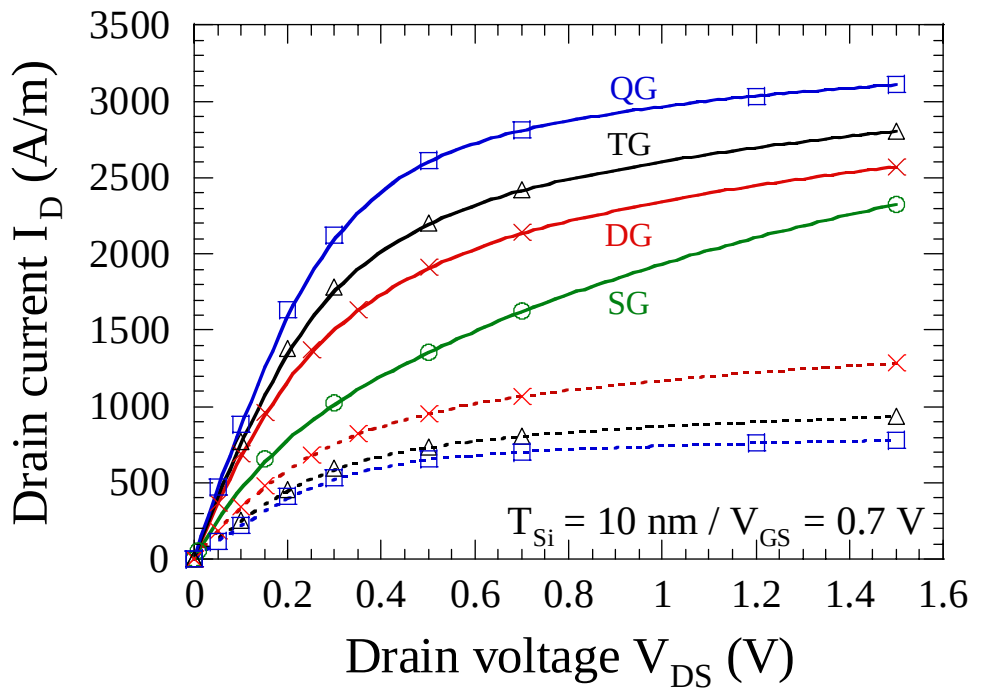

*Figure 2: $I_D$ versus $V_{DS}$ at $V_{GS}$ = 0.7 V in reference devices. In dashed lines, $I_D$ divided by the gate number: by 2 for DG, by 3 for TG and by 4 for QG.*

In reference devices, variations of drain current $I_D$ as a function of gate voltage $V_{GS}$ obtained at low drain voltage $V_{DS}$ (not shown) indicates that $V_T$-values are close: 0.18 V for SG, 0.2 V for DG, and 0.23 V for TG and QG.

Then as shown in dashed lines in Figure 2, the current $I_{on}$ of reference DG, TG and QG are less than 2, 3 and 4 times, respectively, higher than that of the reference SG in which $I_{on}$ is significantly enhanced by strong SCE. So, multiple-gates seem to be less effective in the on-state than a SG of equivalent geometry. It is more problematic to note the following point: the pitch P between two active 3D devices in the multi-fingered architecture has to be small enough to challenge planar devices in a given design area.

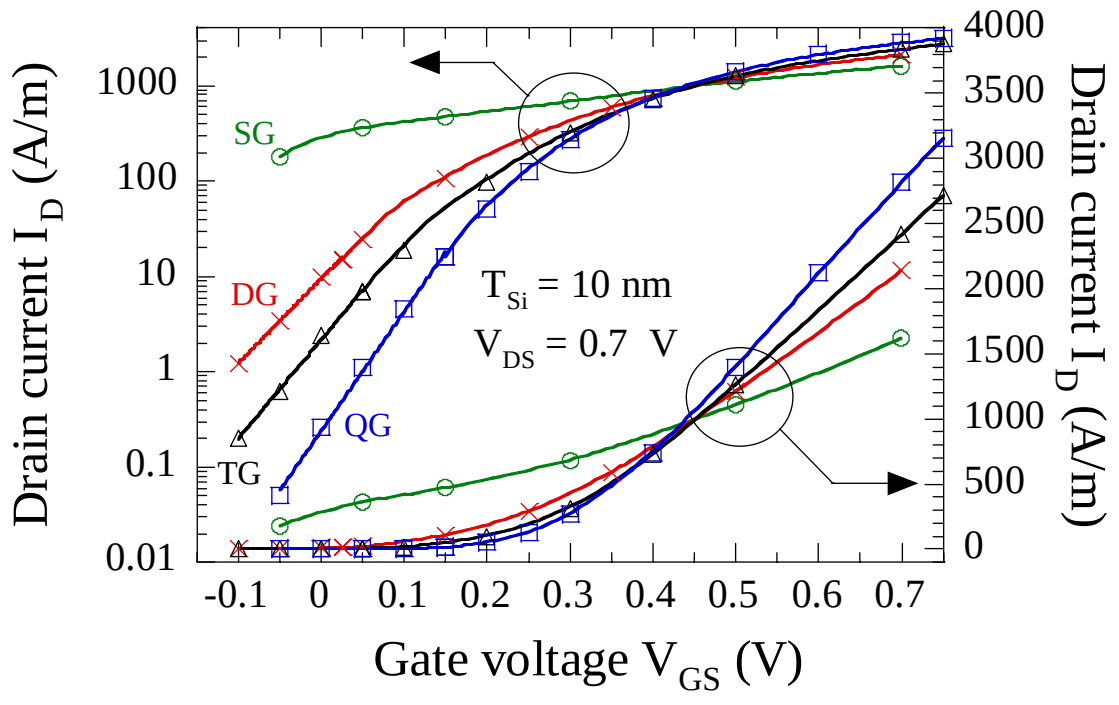

*Figure 3: $I_D(V_{GS})$ at $V_{GS}$ = 0.7 V in reference devices.*

### 2.2 Subthreshold characteristics

As represented in Figure 3, in the reference SG the off-state current $I_{off}$ at $V_{GS}$ = 0 V and $V_{DS} = V_{DD}$ is unacceptably high: 200 A/m. Such poor characteristics are related to a high leakage current at the body/BOX interface (not shown).

The subthreshold characteristics of DG are clearly better than that of the SG but remain far from ITRS recommendations, despite the fact that $L_{ch}$ is equal to $3T_{Si}/2$ [3]. For example, the subthreshold slope S is 110 mV/dec at $V_{DS}$ = 0.7 V while it should not be higher than 80 mV/dec. The conduction band evolution along the DG in the centre of the body ($y = T_{Si}/2$) is plotted in Figure 4 for different bias voltages. This illustrates clearly the drain induced barrier lowering (DIBL). The insert of Figure 4 showing the evolution of the conduction band injection barrier as a function of the bias voltage, highlights that the DIBL is not a linear function of the drain voltage $V_{DS}$. Besides, estimating the evolution of the barrier for $V_{DS}$ varying between 0.05 V and 0.7 V, appears to be sufficient. So, for $V_{GS}$ = 0 V and $V_{DS}$ varying between 0.05 V and 0.7 V, the DIBL is 73 mV i.e. a relative variation of 35%. Just underneath the top gate, the DIBL is weaker: 46 mV (16%). So, the form factor $L_{ch}/T_{Si}$ is in fact too weak to prevent from significant SCE in the centre of the body that is the region furthest from the gates. It also explains the high values of $V_T$ roll-off: $\Delta V_T$ = -135 mV for $V_{DS}$ varying between 0.05 V and 0.7 V.

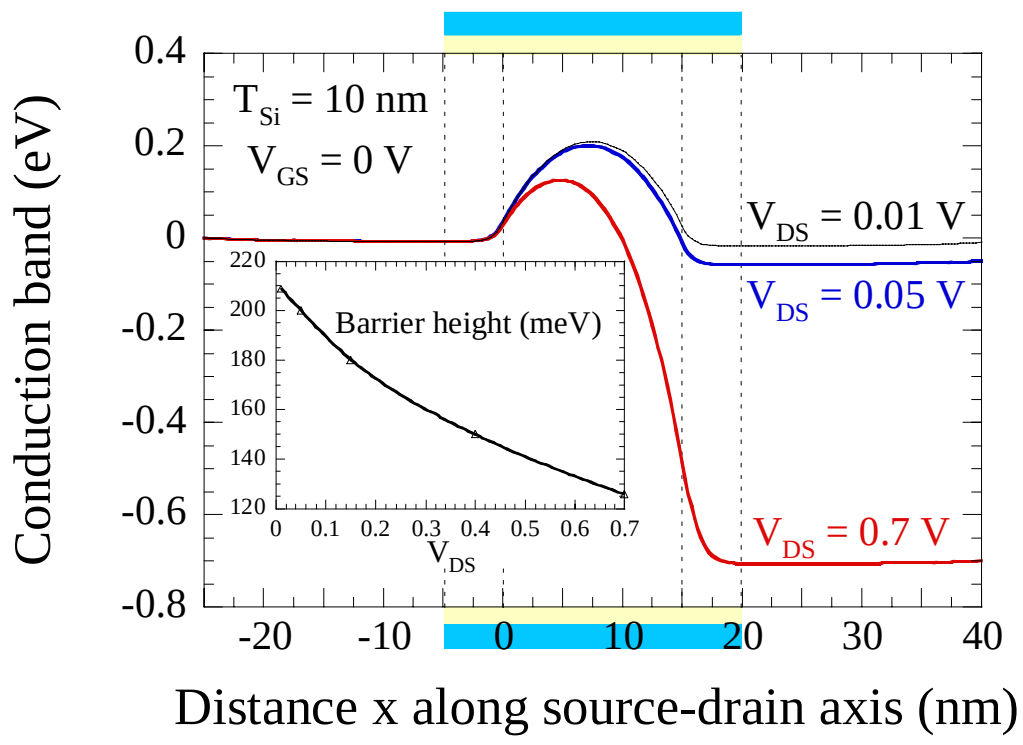

*Figure 4: Conduction band in the reference DG along the x-axis at $y = T_{Si}/2$ and $V_{GS}$ = 0 V. Insert: DIBL as a function of the drain voltage $V_{DS}$.*

In the reference TG and QG, the gate control is improved. For comparison, the electron density n is plotted in Figure 5 and Figure 6 in the reference DG and QG, and SG and TG, respectively, as a function of the distance y between the top and bottom gates, at $x = L_{ch}/2$, z = W/2, low $V_{DS}$ and different $V_{GS}$ values. For $V_{GS}$ varying between 0 V and 0.3 V, n is quite homogenous all along the TG and QG thickness $T_{Si}$, contrary to the cases of the SG and DG where the density near the buried oxide and in the body centre ($n_c$) respectively are much greater than that at $SiO_2$/Si interface ($n_i$). So a significant leakage current takes place near the buried oxide for SG and in the centre of the body for DG which are deficiently controlled by the gates. For higher $V_{GS}$ values, $n_i$ is greater than $n_c$ in both cases but the difference is higher in QG than in DG: $n_i/n_c$ is equal to about 11 in QG and 19 in DG at $V_{GS}$ = 0.7 V.

As a consequence, S and $\Delta V_T$ are reduced in reference TG (resp. 96 mV/dec and 100 mV) and QG (resp. 83 mV/dec and 44 mV). Those values remain however a bit too high for practical applications, despite the fact that $L_{ch}$ is greater than $T_{Si}$ = W [3]. At $V_{GS}$ = 0 V and in the region furthest away from the gates, the DIBL is equal to 76 mV (23%) in TG and 35 mV (12%) in QG.

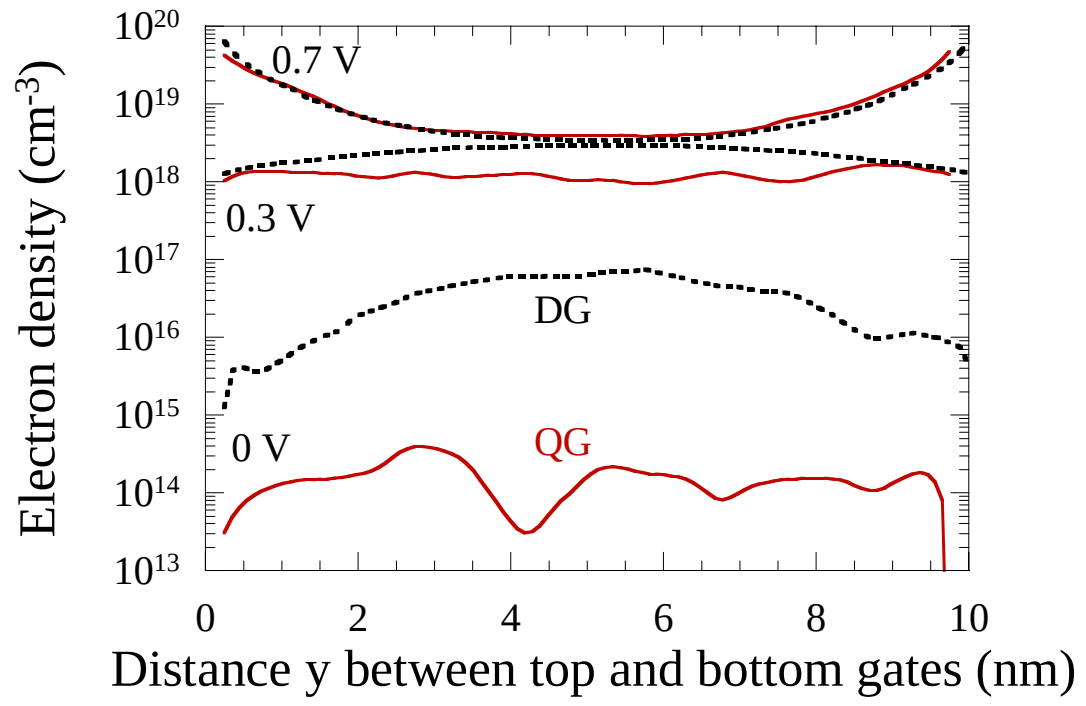

*Figure 5: Electron density versus y in the reference DG and QG at $x = L_{ch}/2$, z = W/2, and low $V_{DS}$.*

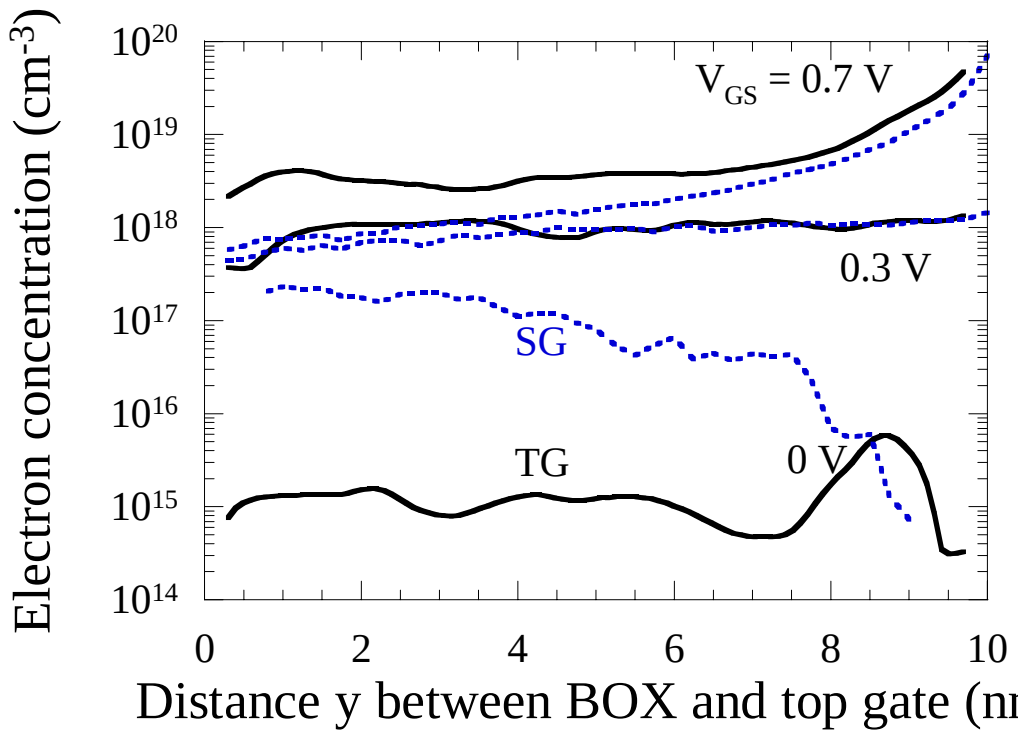

*Figure 6: Electron density versus y in the reference SG and TG at $x = L_{ch}/2$, z = W/2 and low $V_{DS}$.*

### 2.3 Devices with thinner body

Expecting improved device performance by reducing the channel thickness, we have simulated devices with $T_{Si}$ = W = 5 nm. The curves $I_D(V_{DS})$ at $V_{GS}$ = 0.7 V and $I_D(V_{GS})$ at $V_{DS}$ = 0.7 V are drawn in solid lines in Figure 7 and Figure 8 for the thinner SG, DG and QG. Because of the greater $L_{ch}/T_{Si}$ form factor (equal to 3), the electrical characteristics of SG become more acceptable. But they are far to satisfy the ITRS recommendations in the subthreshold regime: S = 127 mV/dec, $\Delta V_T$ = -200 mV. To really control short channel effects in SG, $T_{Si}$ has to be less than $L_{ch}/3$ or $T_{ox}$ must be much thinner.

With thinner Si body, device performances increase once again with the number of the gate: S is for example equal to 80 mV/dec in DG, 69 mV/dec in TG and 64 mV/dec in QG. However, all the performances of the thin DG are satisfactory regarding CMOS application: $I_{on}$ = 1280 A/m, $\Delta V_T$ = -24 mV, $g_D$ = 150 S/m and $g_m$ = 3350 S/m. Moreover, as illustrated in Figure 8 on a logarithmic scale, the DG subthreshold behavior is much closer to that of the TG and QG than in the case of reference devices. Indeed, the gate control on electron density in this device is as good as, or even better than, that in the reference QG. It should be mentioned that, as in reference devices, $I_{on}$ is not proportional to the gate number as shown by comparison between dashed lines and SG characteristic in Figure 7.

The conduction band evolution along the thin DG shows however a significant ohmic drop in the highly doped source region. It corresponds to a source resistance $R_S$ equal to 110 μΩ.m while no contribution of contact resistance has been taken into account. ITRS recommends values less than 140 μΩ.m. Hence, series source/drain resistances $R_{SD}$ become a serious difficulty in such ultrathin devices.

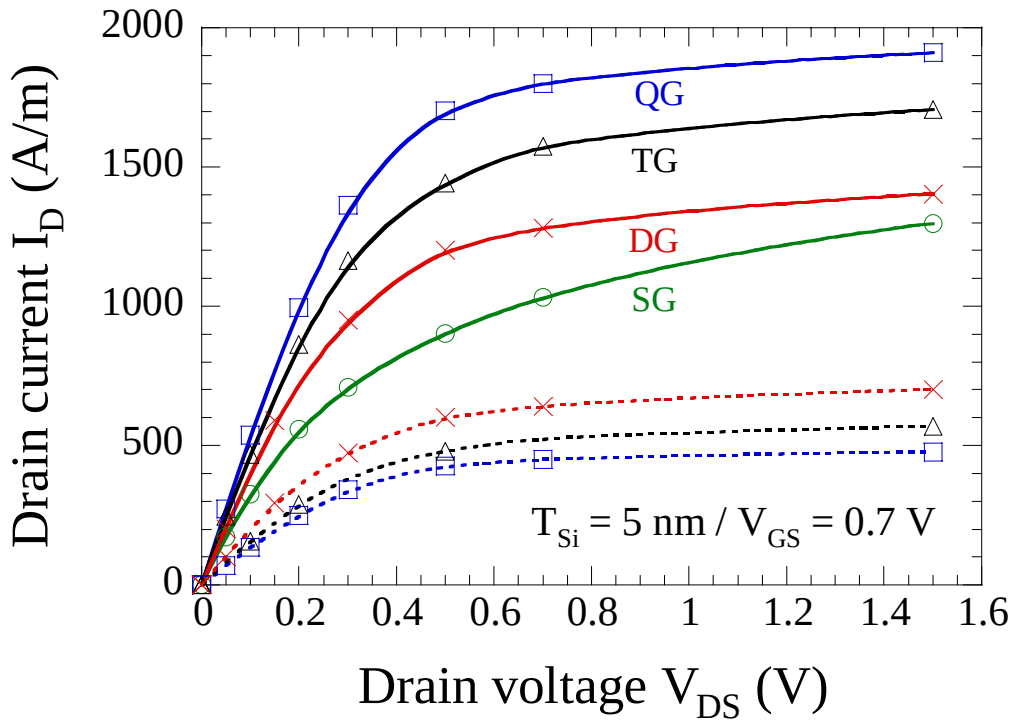

*Figure 7: $I_D$ versus $V_{DS}$ at $V_{GS}$ = 0.7 V in devices with thinner body ($T_{Si}$ = 5 nm). In dashed lines, $I_D$ divided by the gate number.*

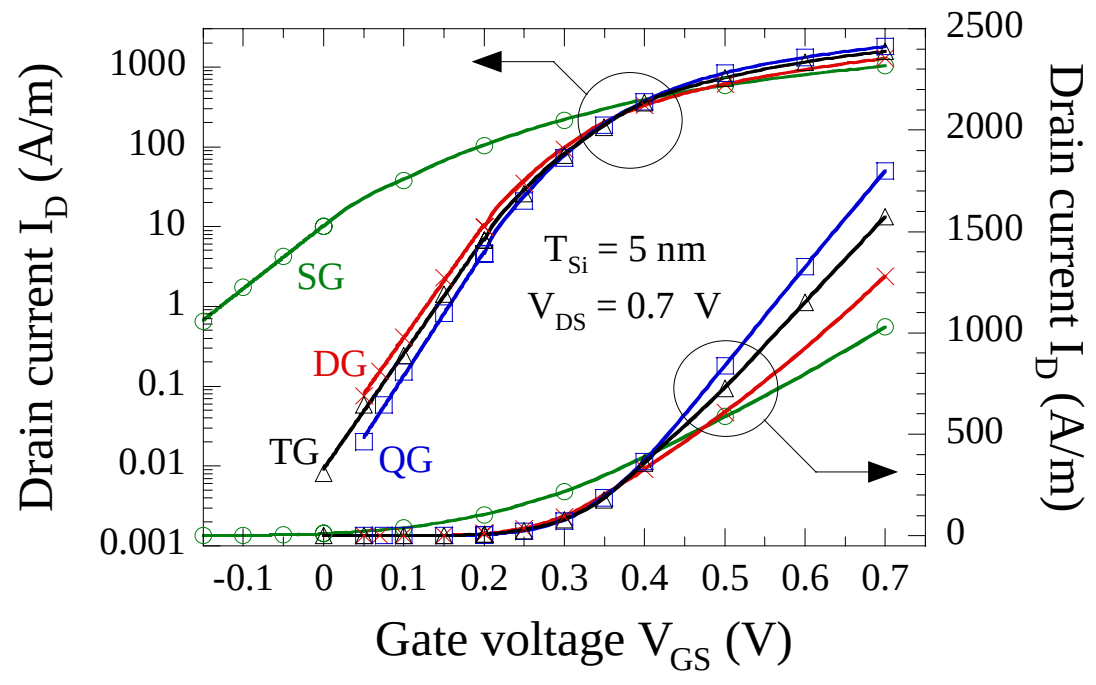

*Figure 8: $I_D(V_{GS})$ at $V_{DS}$ = 0.7 V in thin devices.*

### 2.4 Increase of the effective channel length

We have simulated devices with $T_{Si}$ = 10 nm and longer effective channel: $L_{ch}$ = $L_G$ = 25 nm. To sum up all results, $I_{on}$ is plotted as a function of $I_{off}$ in Figure 9. As expected for the 25 nm-long devices, both $I_{on}$ and short channel effects are smaller, the difference between the subthreshold behavior of the DG and that of the TG and QG is also weaker (S = 70 mV/dec and 67 mV/dec, respectively) than in the reference device. The saturation of the $I_{on}/I_{off}$ ratio as a function of the increase of the gate number (shown in the insert) also illustrates this trend.

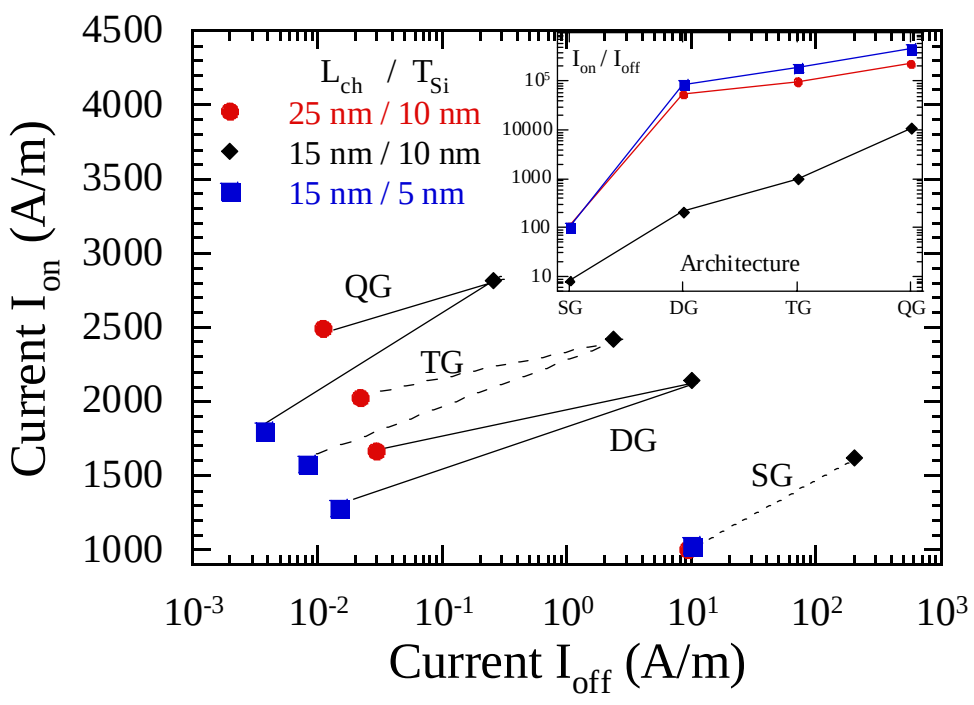

*Figure 9: $I_{on}$ versus $I_{off}$ plot for all simulated devices (lines are only guidelines for the eye). Insert: $I_{on}$ / $I_{off}$ ratio as a function of the architecture.*

### 2.5 Delay time analysis and discussion

Despite a thick gate oxide, the multiple-gate devices TG and QG have a very remarkable off-state behavior. However, the study of $C_G V_{DD}/I_{on}$ delay, plotted in Figure 10 as a function of $I_{off}$ for all studied devices, moderates those conclusions. The intrinsic MOS capacitance $C_G$ is calculated from Monte Carlo results, as described in [6]. $C_G V_{DD}/I_{on}$ investigations show an obvious advantage to SG devices which present, in parallel, catastrophic subthreshold behavior. Indeed, the increase of the gate number induces, of course, an off-state improvement but also a strong rise of $C_G$. A properly designed DG appears to be the better compromise at given $I_{off}$, as also shown by the evolution of $I_{on}/I_{off}$ as a function of the gate number in the insert of Figure 9.

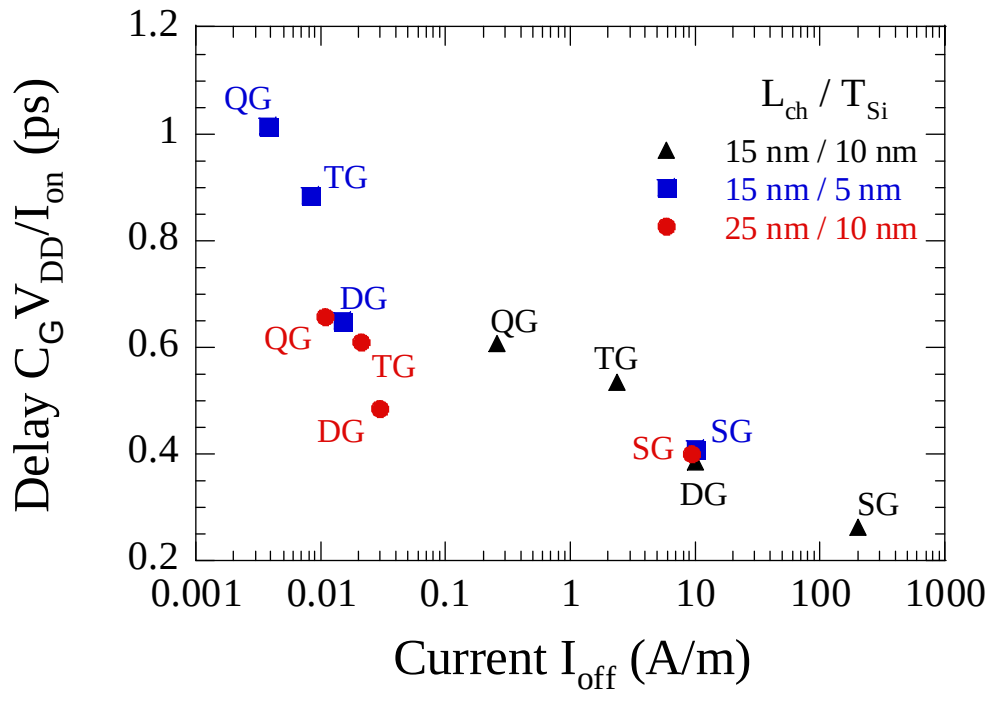

*Figure 10: $C_G V_{DD}/I_{on}$ versus $I_{off}$ plot.*

## 3. Conclusion

We have investigated in detail the electrical characteristics of quasi ballistic multiple-gate MOSFETs. Their efficiency in subthreshold regime compared with SG architectures is obvious. Besides, even if better $I_{on}/I_{off}$ ratios are obtained in TG and QG than in DG, a properly designed planar DG may be a good compromise. The use of thinner body will improve performances if the series resistances increase is controlled.

This work is supported by the French RMNT under project CMOS-D-ALI and by the European Community under Integrated Project NANOCMOS and Network of Excellence SINANO. We also thank M. Vinet and S. Deleonibus for helpful discussions, and E. Grémion for his contribution.

## References

[1] Colinge J-P. Solid-State Electron 2004; 48:897.
[2] Harrison S et al. IEDM 2003.
[3] Chau R et al. Physica E 2003; 19:1.
[4] Fenouillet-Beranger C et al. Solid State Electron 2004, 48:961.
[5] Saint Martin J et al. IEEE Trans Electron Dev 2004; 51:1148.
[6] Dollfus P. J Appl Phys 1997; 82:3911.
[7] International Technology Roadmap for Semiconductors, http://public.itrs.net/.